



\font\bigbf=cmbx10 scaled\magstep2

\font\twelverm=cmr10 scaled 1200    \font\twelvei=cmmi10 scaled 1200
\font\twelvesy=cmsy10 scaled 1200   \font\twelveex=cmex10 scaled 1200
\font\twelvebf=cmbx10 scaled 1200   \font\twelvesl=cmsl10 scaled 1200
\font\twelvett=cmtt10 scaled 1200   \font\twelveit=cmti10 scaled 1200

\skewchar\twelvei='177   \skewchar\twelvesy='60


\def\twelvepoint{\normalbaselineskip=12.4pt
  \abovedisplayskip 12.4pt plus 3pt minus 9pt
  \belowdisplayskip 12.4pt plus 3pt minus 9pt
  \abovedisplayshortskip 0pt plus 3pt
  \belowdisplayshortskip 7.2pt plus 3pt minus 4pt
  \smallskipamount=3.6pt plus1.2pt minus1.2pt
  \medskipamount=7.2pt plus2.4pt minus2.4pt
  \bigskipamount=14.4pt plus4.8pt minus4.8pt
  \def\rm{\fam0\twelverm}          \def\it{\fam\itfam\twelveit}%
  \def\sl{\fam\slfam\twelvesl}     \def\bf{\fam\bffam\twelvebf}%
  \def\mit{\fam 1}                 \def\cal{\fam 2}%
  \def\tt{\twelvett}
  \textfont0=\twelverm   \scriptfont0=\tenrm   \scriptscriptfont0=\sevenrm
  \textfont1=\twelvei    \scriptfont1=\teni    \scriptscriptfont1=\seveni
  \textfont2=\twelvesy   \scriptfont2=\tensy   \scriptscriptfont2=\sevensy
  \textfont3=\twelveex   \scriptfont3=\twelveex 
 \scriptscriptfont3=\twelveex
  \textfont\itfam=\twelveit
  \textfont\slfam=\twelvesl
  \textfont\bffam=\twelvebf \scriptfont\bffam=\tenbf
  \scriptscriptfont\bffam=\sevenbf
  \normalbaselines\rm}



\def\beginlinemode{\endmode
  \begingroup\parskip=0pt 
\obeylines\def\\{\par}\def\endmode{\par\endgroup}}
\def\beginparmode{\endmode
  \begingroup \def\endmode{\par\endgroup}}
\let\endmode=\par
{\obeylines\gdef\
{}}
\def\singlespace{\baselineskip=\normalbaselineskip}
\def\oneandathirdspace{\baselineskip=\normalbaselineskip
  \multiply\baselineskip by 4 \divide\baselineskip by 3}

\def\doublespace{\baselineskip=
\normalbaselineskip \multiply\baselineskip by 2}

\newcount\firstpageno
\firstpageno=1
\footline={\ifnum\pageno<\firstpageno{\hfil}%
\else{\hfil\twelverm\folio\hfil}\fi}
\let\rawfootnote=\footnote              
\def\footnote#1#2{{\rm\singlespace\parindent=0pt\rawfootnote{#1}{#2}}}
\def\raggedcenter{\leftskip=4em plus 12em \rightskip=\leftskip
  \parindent=0pt \parfillskip=0pt \spaceskip=.3333em \xspaceskip=.5em
  \pretolerance=9999 \tolerance=9999
  \hyphenpenalty=9999 \exhyphenpenalty=9999 }
\def\dateline{\rightline{\ifcase\month\or
  January\or February\or March\or April\or May\or June\or
  July\or August\or September\or October\or November\or December\fi
  \space\number\year}}
\def\received{\vskip 3pt plus 0.2fill
 \centerline{\sl (Received\space\ifcase\month\or
  January\or February\or March\or April\or May\or June\or
  July\or August\or September\or October\or November\or December\fi
  \qquad, \number\year)}}


\hsize=6.5truein
\vsize=8.9truein
\voffset=0.0truein
\parskip=\medskipamount
\twelvepoint            
\oneandathirdspace           
\overfullrule=0pt       



\def\title                      
  {\null\vskip 3pt plus 0.2fill
   \beginlinemode \doublespace \raggedcenter \bigbf}

\def\author                     
  {\vskip 3pt plus 0.2fill \beginlinemode
   \singlespace \raggedcenter}

\def\affil                      
  {\vskip 4pt 
\beginlinemode
   \singlespace \raggedcenter \sl}

\def\abstract                   
  {\vskip 3pt plus 0.3fill \beginparmode
   \oneandathirdspace\narrower}

\def\endtitlepage               
  {\endpage                     
   \body}

\def\body                       
  {\beginparmode}               

\def\head#1{                    
  \vskip 0.25truein     
 {\immediate\write16{#1}
   \noindent{\bf{#1}}\par}
   \nobreak\vskip 0.01truein\nobreak}

\def\subhead#1{                 
  \vskip 0.25truein             
  \noindent{{\it {#1}} \par}
   \nobreak\vskip 0.005truein\nobreak}

\def\refto#1{[#1]}           

\def\references                 
  {\subhead{\bf References}         
   \beginparmode
   \frenchspacing \parindent=0pt \leftskip=1truecm
   \doublespace\parskip=8pt plus 3pt
 \everypar{\hangindent=\parindent}}

\gdef\refis#1{\indent\hbox to 0pt{\hss#1.~}}    

\gdef\journal#1, #2, #3, #4#5#6#7{               
    {\sl #1~}{\bf #2}, #3 (#4#5#6#7)}           

\def\refstylenp{                
  \gdef\refto##1{ [##1]}                                
  \gdef\refis##1{\indent\hbox to 0pt{\hss##1)~}}        
  \gdef\journal##1, ##2, ##3, ##4 {                     
     {\sl ##1~}{\bf ##2~}(##3) ##4 }}

\def\refstyleprnp{              
  \gdef\refto##1{ [##1]}                                
  \gdef\refis##1{\indent\hbox to 0pt{\hss##1)~}}        
  \gdef\journal##1, ##2, ##3, 1##4##5##6{               
    {\sl ##1~}{\bf ##2~}(1##4##5##6) ##3}}

\def\prd{\journal Phys. Rev. D, }

\def\jmp{\journal J. Math. Phys., }

\def\cqg{\journal Class. Quantum Grav., }

\def\endreferences{\body}

\def\figurecaptions             
  { \beginparmode
   \subhead{Figure Captions}
}

\def\endpage                    
  {\vfill\eject}

\def\endpaper                   
  {\endmode\vfill\supereject}

\def\hook{\mathbin{\raise2.5pt\hbox{\hbox{{\vbox{\hrule height.4pt 
width6pt depth0pt}}}\vrule height3pt width.4pt depth0pt}\,}}
\def\today{\number\day\ \ifcase\month\or
     January\or February\or March\or April\or May\or June\or
     July\or August\or September\or October\or November\or
     December\space \fi\ \number\year}
\def\date{\noindent{\tt 
     Date typeset: \today\par\bigskip}}
\def\ref#1{Ref. #1}                     
\def\Ref#1{Ref. #1}                     

\def\frac#1#2{{\textstyle{#1 \over #2}}}
\def\half{{\textstyle{ 1\over 2}}}
\def\>{\rangle}
\def\<{\langle}

\def\ie{{\it i.e.,\ }}

\def\sla{\raise.15ex\hbox{$/$}\kern-.57em}
\def\leaderfill{\leaders\hbox to 1em{\hss.\hss}\hfill}
\def\twiddle{\lower.9ex\rlap{$\kern-.1em\scriptstyle\sim$}}
\def\bigtwiddle{\lower1.ex\rlap{$\sim$}}
\def\gtwid{
\mathrel{\raise.3ex\hbox{$>$\kern-.75em\lower1ex\hbox{$\sim$}}}}
\def\ltwid{\mathrel{\raise.3ex\hbox
{$<$\kern-.75em\lower1ex\hbox{$\sim$}}}}
\def\square{\kern1pt\vbox{\hrule height 1.2pt\hbox
{\vrule width 1.2pt\hskip 3pt
   \vbox{\vskip 6pt}\hskip 3pt\vrule width 0.6pt}
\hrule height 0.6pt}\kern1pt}

\def\m@th{\mathsurround=0pt }
\def\leftrightarrowfill{$\m@th \mathord\leftarrow \mkern-6mu
 \cleaders\hbox{$\mkern-2mu \mathord- \mkern-2mu$}\hfill
 \mkern-6mu \mathord\rightarrow$}
\def\overleftrightarrow#1{\vbox{\ialign{##\crcr
     \leftrightarrowfill\crcr\noalign{\kern-1pt\nointerlineskip}
     $\hfil\displaystyle{#1}\hfil$\crcr}}}


\font\titlefont=cmr10 scaled\magstep3

\def\martinstyletitle                      
  {\null\vskip 3pt plus 0.2fill
   \beginlinemode \doublespace \raggedcenter \titlefont}

\font\twelvesc=cmcsc10 scaled 1200

\def\author                     
  {\vskip 3pt plus 0.2fill \beginlinemode
   \singlespace \raggedcenter\twelvesc}


\def\endtitle{\body}
\def\endauthor{\body}
\def\endaffil{\body}

\def\heading                            
  {\vskip 0.5truein plus 0.1truein      
\endheading
   \beginparmode \def\\{\par} \parskip=0pt \singlespace \raggedcenter}

\def\endheading
  {\par\nobreak\vskip 0.25truein\nobreak\beginparmode}

\def\subheading                         
  {\vskip 0.25truein plus 0.1truein     
   \beginlinemode \singlespace \parskip=0pt \def\\{\par}\raggedcenter}

\def\tag#1$${\eqno(#1)$$}

\def\align#1$${\eqalign{#1}$$}

\def\aligntag#1$${\gdef\tag##1\\{&(##1)\cr}\eqalignno{#1\\}$$
  \gdef\tag##1$${\eqno(##1)$$}}

\def\endaligntag{}

\def\overset #1\to#2{{\mathop{#2}\limits^{#1}}}
\def\underset#1\to#2{{\let\next=#1\mathpalette\undersetpalette#2}}
\def\undersetpalette#1#2{\vtop{\baselineskip0pt
\ialign{$\mathsurround=0pt #1\hfil##\hfil$\crcr#2\crcr\next\crcr}}}


\def\ref#1{Ref.~#1}                     
\def\Ref#1{Ref.~#1}                     
\def\[#1]{[\cite{#1}]}
\def\cite#1{[#1]}
\def\(#1){(\call{#1})}
\def\call#1{{#1}}
\def\taghead#1{}
\def\frac#1#2{{#1 \over #2}}
\def\half{{\frac 12}}

\def\12{{1\over2}}

\def\ie{{\it i.e.,\ }}

\def\sla{\raise.15ex\hbox{$/$}\kern-.57em}
\def\leaderfill{\leaders\hbox to 1em{\hss.\hss}\hfill}
\def\twiddle{\lower.9ex\rlap{$\kern-.1em\scriptstyle\sim$}}
\def\bigtwiddle{\lower1.ex\rlap{$\sim$}}
\def\gtwid{\mathrel{\raise.3ex\hbox{$>$
\kern-.75em\lower1ex\hbox{$\sim$}}}}
\def\ltwid{\mathrel{\raise.3ex\hbox{$<$
\kern-.75em\lower1ex\hbox{$\sim$}}}}
\def\square{\kern1pt\vbox{\hrule height 1.2pt\hbox
{\vrule width 1.2pt\hskip 3pt
   \vbox{\vskip 6pt}\hskip 3pt\vrule width 0.6pt}
\hrule height 0.6pt}\kern1pt}
\def\tdot#1{\mathord{\mathop{#1}\limits^{\kern2pt\ldots}}}

\def\pmb#1{\setbox0=\hbox{#1}%
  \kern-.025em\copy0\kern-\wd0
  \kern  .05em\copy0\kern-\wd0
  \kern-.025em\raise.0433em\box0 }

\catcode`@=11
\newcount\tagnumber\tagnumber=0

\immediate\newwrite\eqnfile
\newif\if@qnfile\@qnfilefalse
\def\write@qn#1{}
\def\writenew@qn#1{}
\def\w@rnwrite#1{\write@qn{#1}\message{#1}}
\def\@rrwrite#1{\write@qn{#1}\errmessage{#1}}

\def\taghead#1{\gdef\t@ghead{#1}\global\tagnumber=0}
\def\t@ghead{}

\expandafter\def\csname @qnnum-3\endcsname
  {{\t@ghead\advance\tagnumber by -3\relax\number\tagnumber}}
\expandafter\def\csname @qnnum-2\endcsname
  {{\t@ghead\advance\tagnumber by -2\relax\number\tagnumber}}
\expandafter\def\csname @qnnum-1\endcsname
  {{\t@ghead\advance\tagnumber by -1\relax\number\tagnumber}}
\expandafter\def\csname @qnnum0\endcsname
  {\t@ghead\number\tagnumber}
\expandafter\def\csname @qnnum+1\endcsname
  {{\t@ghead\advance\tagnumber by 1\relax\number\tagnumber}}
\expandafter\def\csname @qnnum+2\endcsname
  {{\t@ghead\advance\tagnumber by 2\relax\number\tagnumber}}
\expandafter\def\csname @qnnum+3\endcsname
  {{\t@ghead\advance\tagnumber by 3\relax\number\tagnumber}}

\def\equationfile{%
  \@qnfiletrue\immediate\openout\eqnfile=\jobname.eqn%
  \def\write@qn##1{\if@qnfile\immediate\write\eqnfile{##1}\fi}
  \def\writenew@qn##1{\if@qnfile\immediate\write\eqnfile
    {\noexpand\tag{##1} = (\t@ghead\number\tagnumber)}\fi}
}

\def\callall#1{\xdef#1##1{#1{\noexpand\call{##1}}}}
\def\call#1{\each@rg\callr@nge{#1}}

\def\each@rg#1#2{{\let\thecsname=#1\expandafter\first@rg#2,\end,}}
\def\first@rg#1,{\thecsname{#1}\apply@rg}
\def\apply@rg#1,{\ifx\end#1\let\next=\relax%
\else,\thecsname{#1}\let\next=\apply@rg\fi\next}

\def\callr@nge#1{\calldor@nge#1-\end-}
\def\callr@ngeat#1\end-{#1}
\def\calldor@nge#1-#2-{\ifx\end#2\@qneatspace#1 %
  \else\calll@@p{#1}{#2}\callr@ngeat\fi}
\def\calll@@p#1#2{\ifnum#1>#2{\@rrwrite
{Equation range #1-#2\space is bad.}
\errhelp{If you call a series of equations by the notation M-N, then M and
N must be integers, and N must be greater than or equal to M.}}\else %
{\count0=#1\count1=
#2\advance\count1 by1\relax\expandafter\@qncall\the\count0,%
  \loop\advance\count0 by1\relax%
    \ifnum\count0<\count1,\expandafter\@qncall\the\count0,%
  \repeat}\fi}

\def\@qneatspace#1#2 {\@qncall#1#2,}
\def\@qncall#1,{\ifunc@lled{#1}{\def\next{#1}\ifx\next\empty\else
  \w@rnwrite{Equation number \noexpand\(>>#1<<) 
has not been defined yet.}
  >>#1<<\fi}\else\csname @qnnum#1\endcsname\fi}

\let\eqnono=\eqno
\def\eqno(#1){\tag#1}
\def\tag#1$${\eqnono(\displayt@g#1 )$$}

\def\aligntag#1\endaligntag
  $${\gdef\tag##1\\{&(##1 )\cr}\eqalignno{#1\\}$$
  \gdef\tag##1$${\eqnono(\displayt@g##1 )$$}}

\def\eqalignno#1{\displ@y \tabskip\centering
  \halign to\displaywidth{\hfil$\displaystyle{##}$\tabskip\z@skip
    &$\displaystyle{{}##}$\hfil\tabskip\centering
    &\llap{$\displayt@gpar##$}\tabskip\z@skip\crcr
    #1\crcr}}

\def\displayt@gpar(#1){(\displayt@g#1 )}

\def\displayt@g#1 {\rm\ifunc@lled{#1}\global\advance\tagnumber by1
        {\def\next{#1}\ifx\next\empty\else\expandafter
        \xdef\csname
 @qnnum#1\endcsname{\t@ghead\number\tagnumber}\fi}%
  \writenew@qn{#1}\t@ghead\number\tagnumber\else
        {\edef\next{\t@ghead\number\tagnumber}%
        \expandafter\ifx\csname @qnnum#1\endcsname\next\else
        \w@rnwrite{Equation \noexpand\tag{#1} is 
a duplicate number.}\fi}%
  \csname @qnnum#1\endcsname\fi}

\def\ifunc@lled#1{\expandafter\ifx\csname @qnnum#1\endcsname\relax}

\let\@qnend=\end\gdef\end{\if@qnfile
\immediate\write16{Equation numbers 
written on []\jobname.EQN.}\fi\@qnend}

\catcode`@=12

\catcode`@=11
\newcount\r@fcount \r@fcount=0
\newcount\r@fcurr
\immediate\newwrite\reffile
\newif\ifr@ffile\r@ffilefalse
\def\w@rnwrite#1{\ifr@ffile\immediate\write\reffile{#1}\fi\message{#1}}

\def\writer@f#1>>{}
\def\referencefile{
  \r@ffiletrue\immediate\openout\reffile=\jobname.ref%
  \def\writer@f##1>>{\ifr@ffile\immediate\write\reffile%
    {\noexpand\refis{##1} = \csname r@fnum##1\endcsname = %
     \expandafter\expandafter\expandafter\strip@t\expandafter%
     \meaning\csname r@ftext
\csname r@fnum##1\endcsname\endcsname}\fi}%
  \def\strip@t##1>>{}}

\def\citeall#1{\xdef#1##1{#1{\noexpand\cite{##1}}}}
\def\cite#1{\each@rg\citer@nge{#1}}	

\def\each@rg#1#2{{\let\thecsname=#1\expandafter\first@rg#2,\end,}}
\def\first@rg#1,{\thecsname{#1}\apply@rg}	
\def\apply@rg#1,{\ifx\end#1\let\next=\relax
\else,\thecsname{#1}\let\next=\apply@rg\fi\next}

\def\citer@nge#1{\citedor@nge#1-\end-}	
\def\citer@ngeat#1\end-{#1}
\def\citedor@nge#1-#2-{\ifx\end#2\r@featspace#1 
  \else\citel@@p{#1}{#2}\citer@ngeat\fi}	
\def\citel@@p#1#2{\ifnum#1>#2{\errmessage{Reference range #1-
#2\space is bad.}%
    \errhelp{If you cite a series of references by the notation M-N, then M 
and
    N must be integers, and N must be greater than or equal to M.}}\else%
 {\count0=#1\count1=#2\advance\count1 
by1\relax\expandafter\r@fcite\the\count0,
  \loop\advance\count0 by1\relax
    \ifnum\count0<\count1,\expandafter\r@fcite\the\count0,%
  \repeat}\fi}

\def\r@featspace#1#2 {\r@fcite#1#2,}	
\def\r@fcite#1,{\ifuncit@d{#1}
    \newr@f{#1}%
    \expandafter\gdef\csname r@ftext\number\r@fcount\endcsname%
                     {\message{Reference #1 to be supplied.}%
                      \writer@f#1>>#1 to be supplied.\par}%
 \fi%
 \csname r@fnum#1\endcsname}
\def\ifuncit@d#1{\expandafter\ifx\csname r@fnum#1\endcsname\relax}%
\def\newr@f#1{\global\advance\r@fcount by1%
    \expandafter\xdef\csname r@fnum#1\endcsname{\number\r@fcount}}

\let\r@fis=\refis			
\def\refis#1#2#3\par{\ifuncit@d{#1}
   \newr@f{#1}%
   \w@rnwrite{Reference #1=\number\r@fcount\space is not cited up to
 now.}\fi%
  \expandafter
\gdef\csname r@ftext\csname r@fnum#1\endcsname\endcsname%
  {\writer@f#1>>#2#3\par}}

\def\ignoreuncited{
   \def\refis##1##2##3\par{\ifuncit@d{##1}%
    \else\expandafter\gdef
\csname r@ftext\csname r@fnum##1\endcsname\endcsname%
     {\writer@f##1>>##2##3\par}\fi}}

\def\r@ferr{\endreferences\errmessage{I was expecting to see
\noexpand\endreferences before now;  I have inserted it here.}}
\let\r@ferences=\references
\def\references{\r@ferences\def\endmode{\r@ferr\par\endgroup}}

\let\endr@ferences=\endreferences
\def\endreferences{\r@fcurr=0
  {\loop\ifnum\r@fcurr<\r@fcount
    \advance\r@fcurr by 
1\relax\expandafter\r@fis\expandafter{\number\r@fcurr}%
    \csname r@ftext\number\r@fcurr\endcsname%
  \repeat}\gdef\r@ferr{}\endr@ferences}


\let\r@fend=\endpaper\gdef\endpaper{\ifr@ffile
\immediate\write16{Cross References written on 
[]\jobname.REF.}\fi\r@fend}

\catcode`@=12

\citeall\refto		
\citeall\ref		%
\citeall\Ref		%

\ignoreuncited
\def\proof{\bigskip\noindent{\bf Proof:}\par}
\line{\hfill March 28, 2006}
\title
Uniqueness of Solutions to the Helically Reduced Wave Equation with Sommerfeld Boundary Conditions
\endtitle

\author
C.~G.~Torre\footnote*{torre@cc.usu.edu}
\endauthor
\affil
Department of Physics
Utah State University
Logan, UT 84322-4415
USA
\endaffil

\abstract

We consider the helical reduction of the wave equation with an arbitrary source on $(n+1)$-dimensional Minkowski space, $n\geq2$. The reduced equation is of mixed elliptic-hyperbolic type on ${\bf R}^n$. We obtain a uniqueness theorem for solutions on a domain consisting of an $n$-dimensional ball $B$ centered on the reduction of the axis of helical symmetry and satisfying ingoing or outgoing Sommerfeld conditions on $\partial B\approx S^{n-1}$.  Non-linear generalizations of such boundary value problems (with $n=3$) arise in the intermediate phase of binary inspiral in general relativity.
\vfill\eject

\body

\head{1. Introduction}

Recent approaches to the quasi-stationary approximation to the intermediate phase of binary inspiral in general relativity have led to the consideration of reductions of the Einstein equations by a helical Killing vector field (see \refto{Price, Klein2004, Friedman2005} and references therein).  To date, model problems have been analyzed consisting of helical reductions of linear and non-linear wave equations in (3+1)-dimensional Minkowski spacetime with various sources using Sommerfeld conditions on a spherical boundary.   These helically-reduced equations have the challenging feature of being of mixed elliptic-hyperbolic type on their $3$-dimensional domain. More precisely, they are elliptic in an inner cylindrical region surrounding the sources and hyperbolic outside this cylindrical region.  There appear to be no general theorems to handle existence and uniqueness of solutions to partial differential equations of mixed type. Results  tend to be specific to individual equations or limited classes of equations, and even then the equations which have been most studied are defined in $2$ dimensions \refto{Rassias1990}. From the investigations of \refto{Price} it appears that the boundary value problem arising from helical reduction of (linear and non-linear) wave equations using Sommerfeld conditions on an exterior boundary is well-posed. Solutions have been constructed and appear to be unique. This is somewhat remarkable since the boundary intersects both the hyperbolic and elliptic domains. In particular, one might not expect a single (Sommerfeld) condition on a closed boundary to enforce uniqueness of solutions \refto{Price}.

Some light was shed on this issue by the work of \refto{CGT2003} where the helical reduction of the $(2+1)$-dimensional wave equation was shown to define a symmetric-positive system on an annular region in ${\bf R}^2$ such that the Sommerfeld boundary value problem was well-posed --- solutions exist and, in particular, are unique. 
Unfortunately, it is not known how to generalize these results (\ie symmetric positivity of the reduced equation) to higher dimensions. Moreover, the helical reduction of the $(2+1)$-dimensional wave equation leads to a boundary value problem on a two-dimensional region with an outer circular boundary which need never intersect the circle of degeneracy of the symbol of the reduced partial differential equation. In higher dimensions, the spherical outer boundary necessarily intersects the ``light cylinder'' where the symbol is degenerate so the boundary conditions {\it must} be imposed both in the elliptic and in the hyperbolic regions.\footnote*{Unless, of course, the boundary is completely contained in the elliptic region, which is not of physical interest and which, in any case, leads to a standard elliptic boundary value problem.} This makes the problem qualitatively different in the physical $(3+1)$ spacetime dimensions (and in higher dimensions). 

Thus it is of interest both from mathematical physics and gravitational physics viewpoints to better understand the nature of boundary value problems arising from helical reduction of wave equations. Here we shall provide a uniqueness theorem for the helical reduction of the $(n+1)$-dimensional wave equation with arbitrary sources and with Sommerfeld boundary conditions. The proof is remarkably elementary and employs an approach used by Protter to study a generalization of the Tricomi problem \refto{Protter1954}. 

\taghead{2.}
\head{2. The helically-reduced wave equation}

We will be considering the helical reduction of the wave equation with an arbitrary source on $(n+1)$-dimensional Minkowski space, with $n\geq2$. The spacetime manifold is $N={\bf R}^{n+1}$ with metric
$$
\eta=-dt\otimes dt + dx\otimes dx + dy\otimes dy + \delta_{ij} dz^i\otimes dz^j,
\tag
$$ 
where Latin indices $i, j=1,2,\dots,n-2$. The wave equation for $\Psi\colon N\to {\bf R}$ with a prescribed source $F\colon N\to {\bf R}$  is
$$
-\Psi_{tt} + \Psi_{xx} + \Psi_{yy} + \delta^{ij} \Psi_{ij} = F.
\tag waveq
$$
Note we use the notation where subscripts on a function indicate partial derivatives. The helical reduction is accomplished by assuming the source and solutions are invariant with respect to the isometry group ($G$) generated by 
$$
K = \partial_t + \Omega(x\partial_y - y\partial_x),\quad \Omega = const.,
\tag
$$
which is equivalent to
$$
{\cal L}_K F = {\cal L}_K\Psi = 0.
\tag inv
$$
 In cylindrical coordinates $(t,\rho,\phi,z^i)$, the metric and Killing vector field are 
$$
\eta = -dt\otimes dt + d\rho \otimes d\rho + \rho^2 d\phi\otimes d\phi + \delta_{ij} dz^i\otimes dz^j,
\tag
$$ 
$$
K= \partial_t + \Omega\partial_\phi,
\tag
$$
the wave equation is
$$
-\Psi_{tt} + {1\over \rho}\partial_\rho(\rho\Psi_\rho) + {1\over \rho^2} \Psi_{\phi\phi} + \delta^{ij}\Psi_{ij} = F,
\tag
$$
and the invariance condition \(inv) is
$$
\Psi_t = - \Omega \Psi_\phi,\quad F_t = - \Omega F_\phi.
\tag inv2
$$
Introducing $\varphi = \phi-\Omega t$, \(inv2) means there exists functions $u$ and $f$ such that
$$
\Psi(t,\rho,\phi,z^i) = u(\rho,\varphi, z^i),\quad F(t,\rho,\phi,z^i) = f(\rho,\varphi, z^i).
\tag
$$
We then get the reduced equation defining helically-invariant solutions to \(waveq)
$$
{1\over \rho}\partial_\rho(\rho u_\rho) + {\chi(\rho)\over \rho^2} u_{\varphi\varphi} +  \delta^{ij}u_{ij} = f,
\tag redeq
$$
where
$$
\chi(\rho) = 1-\Omega^2\rho^2.
\tag
$$

The locus of points where $\chi(\rho)=0$ is the ``light cylinder''. Inside the light cylinder ($\rho<{1\over\Omega}$) eq. \(redeq) is elliptic and outside the light cylinder ($\rho>{1\over\Omega}$) eq. \(redeq) is hyperbolic.

A useful geometric  interpretation of this reduction is as follows. The set of orbits of the group generated by $K$ defines a manifold $M=N/G\approx {\bf R}^n$. The functions $(\rho,\varphi,z^i)$ are $G$-invariant and define cylindrical coordinates on $M$.  In these coordinates the projection $\pi\colon N\to M$ is simply
$$
\pi(t,x,y,z^i) = (\rho,\varphi,z^i),
\tag pi
$$
and satisfies $\pi_*K=0$. The $G$-invariant functions $F$ and $\Psi$ on $N$ correspond to functions $f$ and $u$ on $M$, respectively, via
$$
F = \pi^* f,\quad \Psi = \pi^* u.
\tag
$$
The inverse metric on $N$ is given by
$$
\eta^{\sharp} = -\partial_t\otimes\partial_t + \partial_\rho\otimes\partial_\rho
+ {1\over \rho^2} \partial_\phi\otimes\partial_\phi+ \delta^{ij}\partial_i\otimes\partial_j.
\tag
$$
Being $G$-invariant, $\eta^\sharp$ projects to a tensor field $q$ on $M$. Using \(pi),
$$
q=\pi_*\eta^\sharp =  \partial_\rho\otimes\partial_\rho
+ {\chi(\rho)\over \rho^2} \partial_\varphi\otimes\partial_\varphi+ \delta^{ij}\partial_i\otimes\partial_j.
\tag
$$
This tensor field is well-defined everywhere on $M$, but it does not determine a metric on $M$ because $a$ has no inverse on the light cylinder. 
While the metric on $N$ does not induce a metric on $M$, the metric volume  form $\epsilon$ on $N$ does define a volume form $\nu$ on $M$ as follows. Define
$$
\omega=K\hook \epsilon,
\tag
$$
which satisfies
$$
L_K\omega = 0,\quad K\hook \omega = 0.
\tag
$$
Consequently, $\omega$ is the pull-back by $\pi$ of a volume form $\nu$ on $M$. It is easy to check that
$$
\nu = \rho d\rho\wedge d\varphi\wedge dz^1\wedge\dots\wedge dz^{n-2}.
\tag
$$
The volume form $\nu$ defines a scalar density of weight-1, $\sigma=\rho$, on $M$. 

We will use Greek indices to label tensor fields on $M$.
Introduce a torsion-free derivative operator $\nabla_\alpha$. The reduced equation \(redeq) is equivalent to
$$
{1\over\sigma}\nabla_\alpha(\sigma q^{\alpha\beta} \nabla_\beta u) = f.
\tag eq1
$$
To see this, we first note that, because of the density weights, \(eq1) is in fact independent of the choice of torsion-free derivative $\nabla_\alpha$.  Using the cylindrical coordinate derviative operator, $\nabla_\alpha = \partial_\alpha$, in \(eq1) we obtain \(redeq).
For what follows we re-write \(eq1) as
$$
\nabla_\alpha(h^{\alpha\beta} \nabla_\beta u) 
=\partial_\alpha(h^{\alpha\beta}u_\beta) = \tilde f,
\tag finaleq
$$
where $\tilde f=\sigma f$ is a scalar density of weight one and $h^{\alpha\beta}=\sigma q^{\alpha\beta}$ is a tensor density of weight-1 given by
$$
h^{\rho\rho} =   \rho,\quad  h^{ij} = \rho\delta^{ij},\quad h^{\varphi\varphi} = {1\over \rho}\chi = {1\over \rho} - \Omega^2 \rho.
\tag
$$
 

\taghead{3.}
\head{3. Energy integral}

The key ingredient in our uniqueness theorem is the following generalized energy integral. Fix a domain $B\subset M$ and define
$$
E[u] = \int_B \left[(a u + b^\gamma u_\gamma)\partial_\alpha(h^{\alpha\beta}u_\beta)\right],
\tag
$$
where $a$ and $b^\alpha\partial_\alpha$ are a function and vector field to be specified later.  
The integrand involving $a$ can be written as
$$
au\partial_\alpha(h^{\alpha\beta}u_\beta) = 
\half\partial_\alpha(h^{\alpha\beta}a_\beta) u^2 - a h^{\alpha\beta} u_\alpha u_\beta
+ \partial_\alpha\left[a uh^{\alpha\beta}u_\beta - \half h^{\alpha\beta}a_\beta u^2\right].
\tag
$$
The integrand involving $b^\gamma$ can be written as
$$
b^\gamma u_\gamma\partial_\alpha(h^{\alpha\beta}u_\beta)
=\half \partial_\gamma(h^{\alpha\beta}b^\gamma) u_\alpha u_\beta 
- b^\gamma_{,\alpha} h^{\alpha\beta} u_\gamma u_\beta 
+ \partial_\alpha\left[b^\gamma u_\gamma h^{\alpha\beta} u_\beta
- \half b^\alpha h^{\beta\gamma} u_\gamma u_\beta\right].
\tag
$$
Again, while these expressions use the coordinate derivative, they are in fact independent of the choice of torsion-free derivative operator.
The divergences integrate to the boundary and we have
$$
\eqalign{
E[u] = &\int_B\left\{\half\partial_\alpha(h^{\alpha\beta}a_\beta) u^2 - a h^{\alpha\beta} u_\alpha u_\beta +
\half \partial_\gamma(h^{\alpha\beta}b^\gamma) u_\alpha u_\beta 
- b^\gamma_{,\alpha} h^{\alpha\beta} u_\gamma u_\beta \right\}\cr
&+\int_{\partial B} n_\alpha\left\{(a u + b^\gamma u_\gamma)h^{\alpha\beta}u_\beta - \half h^{\alpha\beta}a_\beta u^2 
- \half b^\alpha h^{\beta\gamma} u_\gamma u_\beta\right\}.
}
\tag finale
$$
If there were a metric on $B$, $n_\alpha$ could be defined in terms of the unit normal to the boundary and the metric-induced volume element of the boundary. Without a metric $n_\alpha$ is still defined, of course, but its definition is necessarily more involved. We give the definition in the Appendix. 

\taghead{4.}
\head{4. Uniqueness theorem}

We are now ready to formulate the boundary value problem of interest. We consider solutions to the equation \(finaleq) on a ball of radius $R$:
$$
B=\{(\rho,\varphi,z^i)| 0\leq\rho^2 + \delta_{ij}z^i z^j \leq R^2\}.
\tag
$$
The boundary $\partial B$ is the sphere $S^{n-1}$ of radius $R$.  Using 
(A.7), we have in spherical coordinates $(r,\theta_1,\dots,\theta_{n-1})$ on $B$ :
$$
n_\alpha dx^\alpha = dr.
\tag
$$
We impose Sommerfeld conditions on $\partial B$. Taking account of the helical reduction they are of the form
$$
{1\over R}({\rho u_\rho + z^i u_i) \pm \Omega \partial_\varphi u} = \tau,\quad {\rm on}\ \partial B
\tag BC
$$
where $\tau\colon S^{n-1}\to {\bf R}$ is some specified function. 

We remark: (i) if the boundary is chosen such that $R>{1\over\Omega}$ the boundary passes through both the elliptic and hyperbolic domains; (ii)  $\tau$ and $\tilde f$ cannot be specified independently; the integral of \(finaleq) over $B$ implies
$$
\int_B \tilde f = \int_{\partial B}\sigma\tau.
\tag
$$

Since \(finaleq) and the boundary conditions \(BC) only involve derivatives of $u$, solutions to these equations can only be unique up to an additive constant. In fact, this is the only freedom in the solution. Our main result is the following.

\proclaim Theorem.
Given $\Omega$, $\tilde f\colon B\to {\bf R}$, and $\tau\colon \partial B\to {\bf R}$, any two solutions to \(finaleq) on $B$ with boundary conditions \(BC)  differ at most by a constant. 

\proof 

Consider the difference of two solutions, $u=u_1-u_2$; $u$ satisfies \(finaleq) and \(BC) with $\tilde f=0$ and $\tau=0$, respectively.  Consequently, $E[u]=0$ for any choices of the function $a$ and vector field $b=b^\alpha\partial_\alpha$. We choose these as
$$
a= -1,\quad b =  {2\over 1-n}\left[\rho\partial_\rho + z^i\partial_i \pm R\Omega\partial_\varphi\right].
\tag
$$
Note that 
$$
b^\alpha u_\alpha = 0,\quad {\rm on}\ \partial B.
\tag
$$
A straightforward computation, using \(BC) with $\tau=0$ in the boundary integral, then gives
$$
\eqalign{
0 &= \int_B\left\{\left({1\over n-1}\right) \sigma\left[(u_\rho^2 + \delta^{ij} u_i u_j) + ({1\over \rho^2} +\Omega^2) u_\varphi^2\right]\right\}\cr
&+ \int_{\partial B} \left({1\over n-1}\right) \sigma{R\over \rho^2}\left\{(z^i z^j + \rho^2\delta^{ij})u_i u_j + (1 + \Omega^2\delta_{ij}z^i z^j)u_\varphi^2 
 \pm 2 R\Omega z^i   u_iu_\varphi\right\}.
}\tag noneg
$$
The volume integrand (in the first integral) is manifestly non-negative for $n\geq 2$. We now show that the boundary integrand (in the second integral) is also non-negative.

We first note that the boundary integrand is invariant under orthogonal transformations of the $z^i$. Thus, given any point $(\rho,\varphi,z^i)$, we can rotate the $z^i$ axes  such that $z^i=(z,0,0,\dots,0)$, where $z^2=\delta_{ij}z^i z^j$. The boundary integrand at the given point is then
$$
\eqalign{
 \left({1\over n-1}\right) &\sigma{R\over \rho^2}\left\{(z^i z^j + \rho^2\delta^{ij})u_i u_j 
 + (1 + \Omega^2\delta_{ij}z^i z^j)u_\varphi^2 
 \pm 2 R\Omega z^i   u_iu_\varphi\right\}\cr
 &= \left({1\over n-1}\right) \sigma{R\over \rho^2}\left\{z^2 u_1^2+ \rho^2\delta^{ij}u_i u_j + (1 + \Omega^2z^2)u_\varphi^2 
 \pm 2 R\Omega z   u_1u_\varphi\right\}\cr
 &\geq 
 \left({1\over n-1}\right) \sigma{R\over \rho^2}\left\{
 u_\varphi^2 + (Ru_1 \pm \Omega z u_\varphi)^2\right\}\cr
 &\geq 0.
 }
 \tag
 $$

Because both integrands in \(noneg) are non-negative they must each vanish. From the volume integrand it follows immediately that
$$
u_\alpha = 0.
\tag
$$
\square

\bigskip\bigskip\noindent
{\bf Acknowledgment}

This work was supported in part by National Science Foundation grant PHY-0244765 to Utah State University.

\bigskip
\taghead{A.}
\head{Appendix: The divergence theorem without a metric}

Consider an $n$-dimensional orientable manifold $M$, a torsion-free derivative operator $\nabla_\alpha$ on $M$, and a vector density of weight one $V^\alpha$. Given $B\subset M$, Stokes theorem implies an identity of the form
$$
\int_B \nabla_\alpha V^\alpha = \int_{\partial B} n_\alpha V^\alpha.
\tag intb
$$
Normally this divergence theorem is proved using a metric on $M$.  However this is not necessary.  Here we shall give a version of the divergence theorem and, in particular, give a formula for $n_\alpha$ without using a metric.

The manifold  $M$, being orientable, comes equipped with a nowhere vanishing $n$-form density of weight minus 1, denoted by $\eta_{\alpha_1\cdots \alpha_n}$, and a totally antisymmetric contravariant tensor density of weight one, $\tilde \eta^{\alpha_1\cdots \alpha_n}$, such that
$$
\tilde\eta^{\alpha_1\cdots \alpha_n} \eta_{\beta_1\cdots \beta_n} = n!\delta^{[\alpha_1}_{\beta_1} \dots \delta^{\alpha_n]}_{\beta_n}.
\tag
$$
Both $\eta_{\alpha_1\cdots \alpha_n}$ and $\tilde\eta^{\alpha_1\cdots \alpha_n}$ are constant for any choice of $\nabla_\mu$. 

The boundary $\partial B$ is an oriented submanifold in $M$ embedded by $i\colon S\to M$, \ie $\partial B = i(S)$.
$S$ is equipped with an $(n-1)$-form density of weight minus one, $\xi_{a_1\dots a_{n-1}}$, and a skew, contravariant rank $(n-1)$ tensor density of weight one, $\tilde\xi^{a_1\dots a_{n-1}}$, satisfying
$$
\tilde\xi^{a_1\dots a_{n-1}}\xi_{b_1\dots b_{n-1}} = (n-1)!
\delta^{[a_1}_{b_1}\dots \delta^{a_{n-1}]}_{b_{n-1}}. 
\tag
$$
(In this Appendix only we use Latin indices to denote tensors on $S$.)

To apply Stokes theorem we define an $(n-1)$-form
$$
\omega_{\alpha_1\dots \alpha_{n-1}} = V^{\beta}\eta_{\beta \alpha_1\cdots \alpha_{n-1}}.
\tag
$$
We then have (using differential form notation)
$$
\eqalign{
\int_B \nabla_\alpha V^\alpha 
&= \int_B d\omega\cr
&=\int_{\partial B} \omega\cr
&=\int_{S} {1\over(n-1)!}\xi^{a_1\dots a_{n-1}}(i^*\omega)_{a_1\dots a_{n-1}}.}
\tag divthm
$$
Now, at points of $\partial B$ we can write
$$
\eqalign{
 \xi^{a_1\dots a_{n-1}}(i^*\omega)_{a_1\dots a_{n-1}}
 &= (i_*\xi)^{\alpha_1\dots \alpha_{n-1}}\omega_{\alpha_1\dots \alpha_{n-1}}\cr
 &=(i_*\xi)^{\alpha_1\dots \alpha_{n-1}}\eta_{\beta\alpha_1\cdots \alpha_{n-1} }V^{\beta}.
 }\tag
 $$
Thus we have
$$
n_\beta = {1\over(n-1)!}(i_*\xi)^{\alpha_1\dots \alpha_{n-1}}\eta_{ \beta \alpha_1\cdots \alpha_{n-1}}.
\tag ndef
$$

An alternative approach to the integral over $B$ in 
\(intb) is to note that it is independent of the choice of $\nabla_\alpha$. If we fix a Riemannian metric $g_{\alpha\beta}$ on $M$, and use the metric compatible derivative operator, we have available the more traditional form of the divergence theorem:
$$
\int_B \nabla_\alpha V^\alpha = \int_{\partial B}\sqrt{\gamma}\, \hat n_\alpha W^\alpha,
\tag divmet
$$
where
$$
W^\alpha = {1\over\sqrt{g}} V^\alpha, 
\tag w
$$
$\hat n_\alpha$ is the outwardly oriented unit normal to $\partial B$, and $\sqrt{\gamma}$ is the induced volume element on $\partial B$.  The result \(divmet) is, of course, equivalent to the manifestly metric independent result \(divthm) above, as can be verified by using the identity
$$
\hat n_\beta = {1\over(n-1)!}{\sqrt{g}\over\sqrt{\gamma}}(i_*\xi)^{\alpha_1\dots \alpha_{n-1}}\eta_{\beta\alpha_1\cdots \alpha_{n-1}}.
\tag
$$

\refis{Protter1954}{M. Protter, \journal Indiana Univ.~Math.~J.~, 3, 435, 1954.}

\refis{Price}{J. Whelan, W. Krivan and R. Price, \cqg 17, 4895, 2000; 
J. Whelan, C. Beetle, W. Landry and R. Price, \cqg 19, 1285, 2002;
Z.~Andrade, {\it et al.,} \prd 70, 064001, 2004;
B.~Bromley, R.~Owen and R.~Price, gr-qc/0502121 (2005);
C.~Beetle, B.~Bromley, and R.~Price, gr-qc/0602027 (2006).}

\refis{Klein2004}{C.~Klein, \prd 70, 124026, 2004.}

\refis{Friedman2005}{J. Friedman, K. Uryu, gr-qc/0510002, (2005).}

\refis{Rassias1990}{J. Rassias, {\sl Lecture Notes on Mixed Type Partial Differential Equations}, (World Scientific, Singapore, 1990).}

\refis{CGT2003}{C.~G.~Torre, \jmp 44, 6223, 2003.}

\references

\endreferences

\bye